\documentclass[aps,prl,reprint,superscriptaddress,floatfix,nobibnotes]{revtex4-2}

\usepackage{graphicx}
\usepackage{amssymb}
\usepackage{amsmath}
\usepackage{color}
\usepackage[colorlinks=true,linkcolor=blue,citecolor=blue,urlcolor=blue]{hyperref}

\usepackage{soul}

\begin{document}

\title{Interaction-Enabled Bulk-Boundary Correspondence without a Global Band Gap} 
\author{Cong Ding}
\affiliation{State Key Laboratory of Quantum Optics Technologies and Devices, Institute of Laser Spectroscopy, Shanxi University, Taiyuan 030006, China}
\affiliation{Collaborative Innovation Center of Extreme Optics, Shanxi University, Taiyuan 030006, China}	

\author{Xuzhen Cao}
\thanks{caoxuzhen1@sxu.edu.cn}
\affiliation{State Key Laboratory of Quantum Optics Technologies and Devices, Institute of Laser Spectroscopy, Shanxi University, Taiyuan 030006, China} 
\affiliation{Collaborative Innovation Center of Extreme Optics, Shanxi University, Taiyuan 030006, China}
\affiliation{School of Computer and Information Technology, Shanxi University, Taiyuan 030006, China}

\author{Zhifu Shi}
\affiliation{Sanli Honors College, School of Physics and Electronics Engineering, Shanxi University, Taiyuan 030006, China} 
\affiliation{Department of Modern Physics, University of Science and Technology of China, Hefei 230026, China}

\author{Yuqing Li}
\thanks{lyqing.2006@163.com}
\affiliation{State Key Laboratory of Quantum Optics Technologies and Devices, Institute of Laser Spectroscopy, Shanxi University, Taiyuan 030006, China} 
\affiliation{Collaborative Innovation Center of Extreme Optics, Shanxi University, Taiyuan 030006, China}		
	  				
\author{Ying Hu}
\thanks{huying@sxu.edu.cn}
\affiliation{State Key Laboratory of Quantum Optics Technologies and Devices, Institute of Laser Spectroscopy, Shanxi University, Taiyuan 030006, China} 
	\affiliation{Collaborative Innovation Center of Extreme Optics, Shanxi University, Taiyuan 030006, China}	

\begin{abstract}
A challenge of growing interest is how to make topology observable when conventional gap-based frameworks no longer suffice. In bulk-edge correspondence, this challenge arises when the global band gap closes and localized boundary modes disappear, even though the bulk topology remains intact. Here we show that interactions unlock a novel emergent boundary response from this spectrally hidden topology without reopening the global band gap. These interaction-induced excitations originate from a bulk-band minimum, yet reconstruct the characteristic boundary channels of a gapped topological parent sharing the same Bloch eigenspaces. We establish this interaction-enabled bulk-boundary correspondence for first- and higher-order topology in one- and two-dimensional bosonic lattices, respectively, and identify a general compatibility principle governing its emergence. Our results extend bulk-boundary correspondence beyond its conventional spectral manifestation, with experimental prospects in cold-atom and photonic platforms.
\end{abstract}

\maketitle
\textcolor{blue}{\textit {Introduction -}} Bulk-boundary correspondence links bulk topology to characteristic, directly observable boundary phenomena~\cite{thouless_quantized_1982,hatsugai_chern_1993,hasan_colloquium_2010,qi_topological_2011,bernevig_topological_2013,graf_bulk-edge_2013}. In conventional gapped topological phases, a global band gap spectrally isolates boundary modes from the bulk continuum. Yet these modes can merge into the continuum and disappear when this spectral protection is lost, even while the underlying Bloch geometry remains intact~\cite{song_observation_2018,jiao_experimentally_2021,pellerin_wave-function_2024,longhi_probing_2018,perez-gonzalez_interplay_2019}. Such apparent breakdown of bulk-boundary correspondence exemplifies a central challenge in topological physics: making topology observable when conventional gap-based frameworks no longer suffice. Recent experiments in lattices with engineered longer-range couplings have brought this challenge into sharp focus~\cite{cheng_critical_2026,lin_critical_2026}, revealing distinct topological structures at band-touching criticality through entanglement observables and boundary-state measurements. Here we instead address a complementary open question: once the global band gap has closed and the corresponding localized boundary modes have disappeared, can the spectrally hidden bulk topology still dictate any emergent boundary structure?

\begin{figure}[tp]
\centering
\includegraphics[width=1\columnwidth]{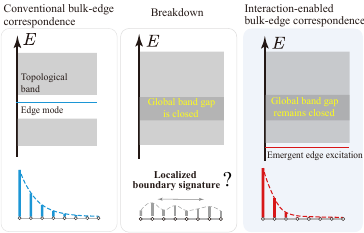}
\caption{\label{Fig1} In conventional gapped topological phases, a global band gap spectrally isolates localized boundary modes tied to bulk topology (left). When gap closure removes these modes but preserves bulk topology (middle), can this hidden topology be made observable at the boundary even \textit{without} reopening the global band gap? Interactions make this possible by inducing a novel boundary excitation (right).
\vspace{-10pt}}
\end{figure}

\begin{figure*}[tp]
	\centering
	\includegraphics[width=1\textwidth]{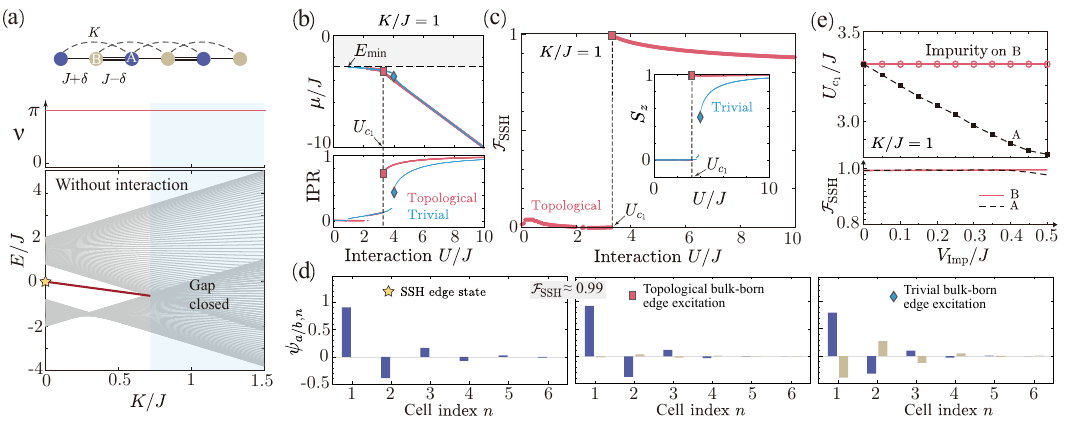}
	\caption{\label{Fig2} Interaction-enabled bulk-edge correspondence of spectrally hidden topology.
(a) Zak phase (top) and open-boundary spectrum (bottom) of the noninteracting extended SSH chain versus NNN hopping $K/J$, for $\delta/J=-0.4$. The shaded region marks closure of the global band gap, where the conventional edge mode (red) disappears while the Zak phase remains unchanged. (b)-(e) Interacting stationary solutions of Eq.~(\ref{eq:NSSH}) for $N=65$ unit cells at $K/J=1$, obtained by boundary-seeded iteration. The global gap in the Bloch bands remains closed. (b) Bifurcation diagram (top): a branch originating from the bulk-band minimum $E_{\rm min}$ undergoes a localization bifurcation at $U_{\rm c1}$, giving rise to a pair of emergent, bulk-born edge excitations related by inversion symmetry; the IPR (bottom) marks their onset of localization. (c) Fidelity $\mathcal F_{\rm SSH}$ of the stationary solution in the topological lattice with the edge mode of the gapped SSH parent ($K=0$), showing near-unit overlap at $U_{\rm c1}$; inset, sublattice polarization $S_z$. In (b) and (c), red (blue) denotes the topological $\delta/J=-0.4$ (trivial $\delta/J=0.4$) lattice.
(d) Spatial profiles of the parent SSH edge mode (left) and emergent edge excitations in the topological (middle) and trivial (right) lattices at their respective localization thresholds. The emergent excitations are born anew from the bulk-band minimum. Only the topological case reconstructs the parent edge channel.
(e) Sublattice-dependent impurity response in the topological lattice: $U_{\rm c1}$ versus impurity strength $V_{\rm imp}$ (top), and $\mathcal F_{\rm SSH}$ at the corresponding threshold (bottom)~\cite{Supple}. Red solid (black dashed) curves correspond to an impurity on the $B$ site of unit cell $n=1$ ($A$ site of unit cell $n=2$).\vspace{-5pt}}
\end{figure*} 

We show that interactions unlock such a boundary signature without reopening the global gap in the underlying Bloch bands. It takes the form of a distinct class of interaction-induced boundary excitations, which are born anew from a bulk-band minimum yet reconstructing the characteristic channels of a gapped topological parent sharing the same Bloch eigenspaces. We systematically demonstrate this interaction-enabled bulk-boundary correspondence for first- and higher-order topology in one- and two-dimensional bosonic lattices, respectively. We also determine a general compatibility principle by which topology, interactions, and spin degrees of freedom conspire to produce this correspondence. Our results establish a qualitatively novel regime of bulk-boundary correspondence, where bulk topology governs emergent boundary structure despite its conventional localized boundary modes have disappeared and the global band gap remains
closed. This mechanism is distinct from interaction-based scenarios built on pre-existing boundary modes or a globally gapped Bloch spectrum~\cite{kirsch_nonlinear_2021,mukherjee_observation_2021,hu_nonlinear_2021,kartashov_observation_2022,pernet_gap_2022,huang_observation_2025,du_observation_2025,maczewsky_nonlinearity-induced_2020,sone_nonlinearity-induced_2024}. Our setup is experimentally feasible in programmable platforms from atomic lattice~\cite{schafer_tools_2020,xie_topological_2020,wang_observation_2022,li_observation_2023,agrawal_two-dimensional_2024} to polaritonic~\cite{st-jean_lasing_2017,solnyshkov_microcavity_2021,pernet_gap_2022} and photonic systems~\cite{bisianov_stability_2019,guo_weakly_2020,mukherjee_observation_2020,maczewsky_nonlinearity-induced_2020,mukherjee_observation_2021,kirsch_nonlinear_2021,hu_nonlinear_2021,pernet_gap_2022,kartashov_observation_2022}.  

\textcolor{blue}{\textit {Bulk topology without localized boundary modes -}} To illustrate how the bulk topology becomes hidden, we begin with the generalized SSH model with next-nearest-neighbor (NNN) coupling [Fig.~\ref{Fig2}(a)]. Its Bloch Hamiltonian reads~\cite{li_topological_2014}
\begin{equation}
h(k)=(2K\cos k) \mathbb I+(J_1+J_2\cos k)\sigma_x+J_2\sin k\sigma_y,\label{hk}
\end{equation}
where $J_{1(2)}=J\pm \delta>0$ is the intracell (intercell) nearest-neighbor coupling rate, $K$ is the NNN hopping, $\mathbb I$ is identity matrix and $\sigma_{x,y}$ is standard Pauli matrix. For $K=0$, this reduces to standard SSH model with chiral symmetry, where the topological ($\delta<0$) and trivial ($\delta>0$) phases are separated by a direct gap closing at the band-inversion momentum $k_{\rm top}=\pi$. Its topological phase hosts zero-energy edge states with sublattice polarization [c.f. Fig.~\ref{Fig2}(d)] 
\begin{equation}\label{S}
S_z=\frac{\sum_n(|\psi_{a,n}|^2-|\psi_{b,n}|^2)}{\sum_n(|\psi_{a,n}|^2+|\psi_{b,n}|^2)}=\pm 1,
\end{equation}
with amplitude $\psi_{a/b,n}$ on sublattice $A/B$ of unit cell $n$. 

\begin{figure}[tp]
\centering
\includegraphics[width=0.85\columnwidth]{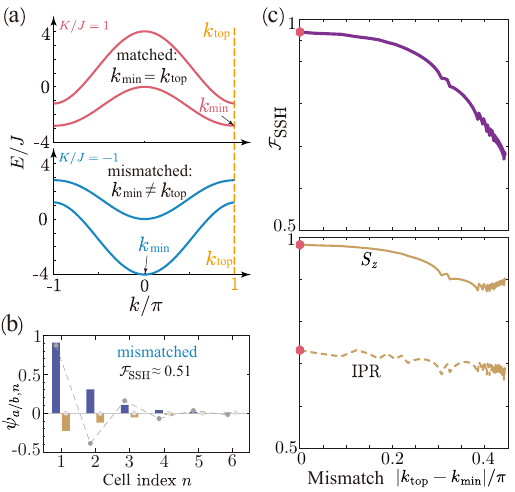}
\caption{\label{Fig3} Momentum matching condition. Results are for the topological extended SSH chain with $\delta/J=-0.4$. 
(a) Energy bands for the matched case $K/J=1$ (top; $k_{\rm top}=k_{\min}=\pi$) and mismatched case $K/J=-1$ (bottom; $k_{\rm top}=\pi\neq k_{\min}=0$). Bloch vectors remain unchanged and the global band gap is closed in both cases. (b) Edge excitation in the mismatched case. Gray curve: $K=0$ SSH edge profile for reference. (c) $\mathcal F_{\rm SSH}$ and $S_z$ decrease with the mismatch $\Delta k=|k_{\rm top}-k_{\min}|$, despite slower change in IPR. The continuous tuning of $\Delta k$ is implemented via additional long-range coupling~\cite{Supple}.\vspace{-10pt}}
\end{figure}

A finite NNN coupling $K\neq 0$ breaks chiral symmetry and strongly reshapes the spectrum [Fig.~\ref{Fig2}(a)]
\begin{equation}
E_\pm(k)=2K\cos k \pm \sqrt{J_1^2+J_2^2+2J_1J_2\cos k}.
\end{equation}
For $2|K|>J-|\delta|$, the global band gap becomes indirect, with the upper-band minimum and lower-band maximum occurring at different $k$ [c.f.~Fig.~\ref{Fig3}(a)]. Crucially, since the NNN term is proportional to the identity, it leaves the Bloch functions $u_\pm(k)$ unchanged, so that the band projectors maintain
\begin{equation}\label{Pk}
P_\pm(k;K)=P_\pm(k;0).
\end{equation}
The Zak phase is thus independent of $K$, with inversion symmetry enforcing quantized values $\nu=\pi$ for $\delta<0$ and $\nu=0$ for $\delta>0$. As long as the indirect gap is open, the topological phase hosts in-gap edge modes. When $2|K|>J+|\delta|$, however, the indirect gap closes without a direct band touching. Consequently, even though the bulk maintains $\nu=\pi$ for $\delta<0$, the would-be topological edge modes vanish [Fig.~\ref{Fig2}(a) blue region]. The usual bulk-edge correspondence thus fails.

\textcolor{blue}{\textit {Interaction-enabled bulk-edge correspondence -}} We now show that adding interactions can induce a new boundary signature from the preserved Bloch eigenspace, \textit{without} reopening the global band gap.  Consider the bosonic extended SSH chain with onsite attractive interactions under open boundaries. Within mean-field theory, the relevant discretized Gross-Pitaevskii (GP) equation ($\hbar\equiv 1$) reads
\begin{eqnarray}\label{eq:NSSH}
\!\!\!\!\!\mu{\psi}_{a,n}&=&J_1{\psi}_{b,n} + J_2\psi_{b,n-1}+K\psi_{a,n-1}+ K\psi_{a,n+1}\nonumber\\
&&-U|\psi_{a,n}|^2\psi_{a,n},\nonumber\\
\!\!\!\!\!\mu{\psi}_{b,n}&=&J_1\psi_{a,n}+J_2\psi_{a,n+1}+K\psi_{b,n-1}+K\psi_{b,n+1}\nonumber\\
&&-U|\psi_{b,n}|^2\psi_{b,n},
\end{eqnarray}
with $U>0$, chemical potential $\mu$ and $\sum_{\alpha,n}|\psi_{\alpha,n}|^2=1$. Such attractive interactions have been realized in 
ultracold atoms~\cite{inouye_observation_1998,cornish_stable_2000,chin_feshbach_2010}. We henceforth focus on the gap-closed regime $2|K|>J+|\delta|$ and $\delta\neq 0$ [c.f. Fig.~\ref{Fig2}(a)].

For topological lattice $\delta<0$, we obtain boundary-seeded stationary solutions of the GP equation by self-consistent iteration~\cite{lumer_self-localized_2013,leykam_edge_2016} (see Supplementary Materials~\cite{Supple}); their inverse participation ratio (IPR), $\textrm{IPR}=\sum_{\alpha=a,b}\sum_n|\psi_{\alpha n}|^4/(\sum_{\alpha,n}|\psi_{\alpha n}|^2)^2$, and spin-polarization $S_z$ [Eq.~(\ref{S})] are shown in Figs.~\ref{Fig2}(b) and (c) as a function of $U$. With increasing $U$, a nonlinear branch bifurcates from the global bulk-band minimum $E_{\rm min}$ at $k_{\min}=\pi$. The branch is initially extended and unpolarized~\cite{Supple}, with $\textrm{IPR}\sim 0$ and $S_z\sim 0$. At a threshold $U_{\rm c1}$, it bifurcates into an inversion-symmetry-related pair of localized edge excitations, with the left (right) state predominantly occupying $A(B)$ sublattice, $S_z\simeq\pm1$.

Most strikingly, these emergent edge excitations $\varphi_{\rm s}$ at $U_{\rm c1}$ have near-unit fidelity with the zero-energy linear edge modes $\varphi_{\rm SSH}$ of the chiral SSH model [Fig.~\ref{Fig2}(c)], 
\begin{equation}
\mathcal F_{\rm SSH}=
|\langle\varphi_{\rm SSH}|\varphi_{\rm s}\rangle|^2\approx1.
\end{equation}
This indicates the nonlinear branch is steered toward the characteristic edge channel of gapped topological parent. By contrast, in the trivial lattice ($\delta>0$), the emergent edge excitations are also localized [Figs.~\ref{Fig2}(b)-(c) blue curve], but they apparently lack topological structure [Fig.~\ref{Fig2}(c) inset, Fig.~\ref{Fig2}(d)]. Hence edge self-localization alone does not select parent edge channel. 

The reconstruction entails a characteristic sublattice-dependent response absent in trivial regime [Fig.~\ref{Fig2}(e)]~\cite{Supple}. An impurity near the left edge on the $B$ sublattice has little effect on $U_{\rm c1}$, whereas an otherwise identical impurity on the $A$ sublattice strongly shifts $U_{\rm c1}$. Nevertheless, the bulk-born edge excitation at $U_{\rm c1}$ robustly maintains $\mathcal F_{\rm SSH}\simeq1$. Moreover, the robustness against coupling disorder is demonstrated in~\cite{Supple}. 

 \begin{figure*}[tp]
\centering
\includegraphics[width=0.9\textwidth]{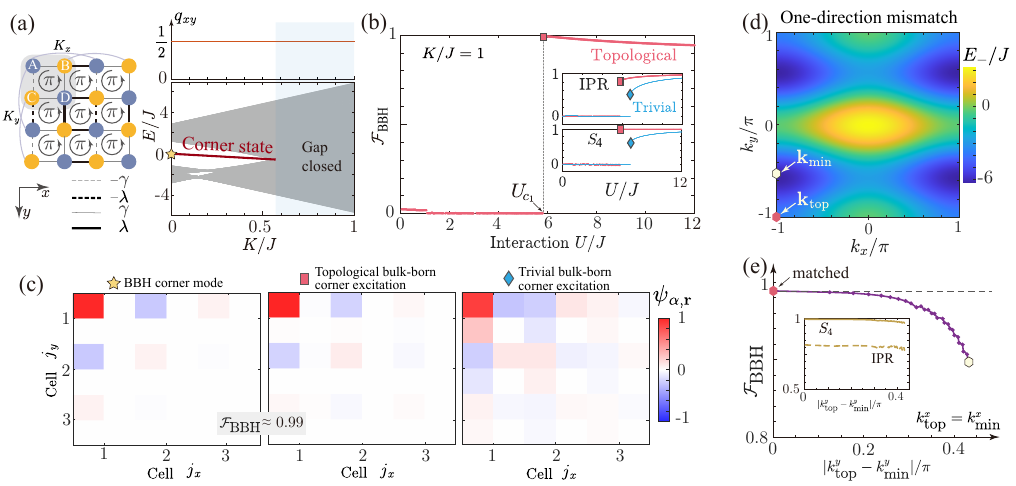}
\caption{\label{Fig4} Interaction-enabled corner correspondence of hidden quadrupole topology without reopening the global band gap. Schematics: extended BBH lattice where $\gamma=J+\delta$, $\lambda=J-\delta$ and a $\pi$ flux threads each plaquette. (a) Quadrupole moment (top) and open-boundary linear spectrum (bottom) versus $K_x=K_y=K$, for $\delta/J=-0.6$ in nontrivial quadrupole regime. (b)-(e) Corner-seeded solutions of the four-component GP equation at $K/J=1$, obtained for 2D lattice with $20\times 20$ unit cells under open boundaries. (b) Fidelity $\mathcal F_{\rm BBH}$ with the $K=0$ parent corner mode versus $U$; inset, IPR and four-site polarization $S_4$. Red and blue correspond to $\delta/J=-0.6$ and $0.6$, respectively. (c) Profile of the parent BBH corner mode, the interaction induced topological and trivial corner excitations. (d) One-direction momentum mismatch, with $k_{\min,x}=k_{{\rm top},x}$ and $k_{\min,y}\ne k_{{\rm top},y}$. (e) $\mathcal F_{\rm BBH}$ as a function of the mismatch in the $y$ direction. \vspace{-10pt} }
\end{figure*}  

\textcolor{blue}{\textit {Compatibility principle -}} To gain insights into the origin of this interaction-enabled bulk-edge correspondence, we first recall how, in a noninteracting gapped phase, the Bloch topology constrains an edge mode - its spatial carrier tied to the band-inversion momentum $k_{\rm top}$ and its intracell texture constrained by protecting symmetry. The edge state (with in-gap energy $E_0$) is an evanescent superposition of analytically continued Bloch solutions~\cite{kohn_analytic_1959,heine_general_1963,mong_edge_2011}, $\psi_n=\sum_j c_j\beta_j^n u_j$, with $[h(\beta_j)-E_0]u_j=0$, $|\beta_j|<1$ and coefficients $c_j$. Here, $e^{ik}\to\beta$ and $u_j$ is the intracell spinor. Since the root associated with the topological edge channel emerges from $e^{ik_{\rm top}}$ at the band inversion, its spatial carrier is tied to $k_{\rm top}$. For chiral SSH parent ($K=0$), the zero-energy left-edge mode is sublattice polarized, $\psi_{b,n}=0$, and its remaining component obeys the boundary kernel, $Q(\beta)\equiv J_1+J_2\beta=0$. Its root yields $\psi_{a,n}\propto (-J_1/J_2)^{n-1}$ that decays only for $J_1<J_2$ and carries $k_{\rm top}=\pi$. Although a sufficiently strong $K$ closes the spectral gap and removes the edge mode, the Bloch eigenspaces and the parent kernel $Q(\beta)$ are preserved. 

Remarkably, interactions provide a route to this retained kernel-defined edge channel template. The reconstruction requires two compatibility conditions. First, a nonlinear branch entering the semi-infinite gap below the linear spectrum is connected, in the weak-amplitude limit, to the global band minimum $E_{\min}$. It therefore inherits the carrier momentum $k_{\min}$ and intracell spinor at that minimum, $\psi_n\sim A_ne^{ik_{\min}n}u_-(k_{\min})$. The branch can access the parent edge template only when its carrier $k_{\min}$ matches the band-inversion momentum,
\begin{equation}\label{match}
{\bf k}_{\min}={\bf k}_{\rm top}.
\end{equation}
For the 1D lattice in Fig.~\ref{Fig2}(b), this gives $k_{\min}=k_{\rm top}=\pi$ [see also Fig.~\ref{Fig3}(a)]. Second, interactions must favor the appropriate intracell texture. In the extended SSH model, $u_-(k_{\rm min})$ is approximately a $\sigma_x$ eigenstate, so the nonlinear branch emerging from $E_{\rm min}$ is initially nearly unpolarized [Fig.~\ref{Fig2}(c)]. Increasing onsite attraction favors a spin-polarized texture, as the interaction energy 
\begin{equation}\nonumber
E_{\rm int}=-\frac{U}{4}\sum_n
\left(\rho_n^2+(s^z_{n})^2\right),
\end{equation}
is lowered by increasing $|s_n^z|=\left||\psi_{a,n}|^2-|\psi_{b,n}|^2\right|$ at fixed $\rho_n=|\psi_{a,n}|^2+|\psi_{b,n}|^2$. At the bifurcation $U_{\rm c1}$, the bulk-connected branch develops inversion-related localized edge excitations. For the $A$-polarized left-edge solution, $\psi_{b,n}\simeq0$, and the $B$ component of Eq.~(\ref{eq:NSSH}) is driven toward $J_1\psi_{a,n}+J_2\psi_{a,n+1}=0$ - the parent SSH kernel. Thus momentum matching supplies the required carrier phase, while interaction-induced polarization exposes the parent kernel. When both conditions are met, they reinforce one another and steer the bulk-born branch toward the self-organized topological edge excitation [Fig.~\ref{Fig2}(d)]. By contrast, no such channel is present in the trivial lattice $\delta>0$. Consequently, although edge-localized excitation still emerges, it is trivial by nature [Fig.~\ref{Fig2}(d)]. 

Figure~\ref{Fig3} shows momentum-matching ~(\ref{match}) is critical for topological reconstruction. Reversing $K\to-K$ from the matched case in Fig.~\ref{Fig2}(b) shifts the band minimum from $k_{\min}=\pi$ to $k_{\min}=0$, while leaving the Bloch eigenspaces, the $k_{\rm top}=\pi$ and the closed gap unchanged [Fig.~\ref{Fig3}(a)]. The resulting edge-localized excitation no longer reconstructs the SSH edge channel [Fig.~\ref{Fig3}(b)]. Fig~\ref{Fig3}(c) shows that the fidelity $\mathcal F_{\rm SSH}$ and $S_z$ decrease as $\Delta k=|k_{\rm top}-k_{\min}|$ increases. Remarkably, even in the $K=0$ SSH limit, the GP equation (\ref{eq:NSSH}) cannot generate a bulk-minimum-born branch that reconstructs its own topological linear edge channel, since $k_{\min}=0\neq k_{\rm top}=\pi$ [Supplemental Fig.~5(a)]. 

\textcolor{blue}{\textit{Interaction-enabled higher-order correspondence -}} 
This interaction-enabled bulk-boundary correspondence extends to higher-order topology, where corner-channel reconstruction requires simultaneous compatibility along multiple spatial directions. Consider an extended Benalcazar-Bernevig-Hughes (BBH) quadrupole lattice~\cite{benalcazar_quantized_2017,benalcazar_electric_2017} in Fig.~\ref{Fig4}(a), whose Bloch Hamiltonian is~\cite{Supple}
\begin{eqnarray}\label{eq:H2ndorder}
\!\!\!\!\!\!\!\!h({\bf k}) &=&\! (2K_x\cos k_x+2K_y\cos k_y)\Gamma_0+(\gamma+ \lambda \cos k_x) \Gamma_4 \nonumber\\
 &-&\!
  (\gamma+\lambda\cos k_y) \Gamma_2\!+\!(\lambda \sin k_x)\Gamma_3\!- \!(\lambda \sin k_y) \Gamma_1. 
\end{eqnarray}
Its doubly degenerate bands are $E_{\pm}({\bf k})=2K_x\cos k_x+2K_y\cos k_y\pm\sqrt{2(\gamma^2+\lambda^2)+2\gamma\lambda(\cos k_x+\cos k_y)}$. The quantized bulk quadrupole moment is nontrivial  $q_{xy}=1/2$ for $\gamma<\lambda$ and trivial $q_{xy}=0$ for $\gamma>\lambda$, independent of $K_{x,y}$. The topological phase transition occurs through the direct gap closing at $\mathbf k_{\rm top}=(\pi,\pi)$. When $K_{x,y}=0$, the topological phase supports four symmetry-related corner modes, e.g., the top-left corner mode [Fig.~\ref{Fig4}(c)], $\varphi_{\rm BBH}({\bf r})\propto\left(-\frac{\gamma}{\lambda}\right)^{j_x+j_y-2}(1,0,0,0)^T$, exhibits single-orbital localization characterized by four-site polarization
\begin{equation}\label{S4}
S_4=
\frac{\sum_{\mathbf r}
(n_{A,\mathbf r}-n_{B,\mathbf r}-n_{C,\mathbf r}+n_{D,\mathbf r})}{\sum_{\mathbf r,\alpha}n_{\alpha,\mathbf r}}=1,
\end{equation}
with $n_{\alpha,\mathbf r}=|\psi_{\alpha,\mathbf r}|^2$ ($\alpha\equiv A,B,C,D$) in unit cell $\mathbf r\equiv (j_x,j_y)$. Increasing $|K_{x,y}|$ closes the spectral gap without changing the Bloch eigenspaces. The quadrupole topology therefore remains encoded in the bulk, while its localized corner modes disappear.

In the gap-closed quadrupole lattice [Fig.~\ref{Fig4}(a)], the corresponding four-component GP equation with the same onsite interaction supports a nonlinear branch connected to the global bulk minimum $E_{\rm min}$~\cite{Supple}. At $U_{\rm c1}$, this branch undergoes a localization bifurcation into four symmetry-related corner excitations; we focus on the top-left solution $\varphi_{\rm s}$. When $\gamma<\lambda$ and $\mathbf{k}_{\min}=\mathbf{k}_{\rm top}=(\pi,\pi)$, the emergent excitation reconstructs the $K=0$ parent corner channel $\varphi_{\rm BBH}$, with $S_4\simeq1$ and $\mathcal F_{\rm BBH}=|\langle\varphi_{\rm BBH}|\varphi_{\rm s}\rangle|^2\simeq1$ [Figs.~\ref{Fig4}(b)-(c)]. Thus the same bulk-born mechanism produces a higher-order boundary manifestation after the original corner modes have vanished.

Interaction-selected corner texture has a simple energetic origin:
\begin{equation}\label{E4}
E_{\rm int}=-\frac{U}{8}\sum_{\mathbf r}
\left(\rho_{\mathbf r}^{\,2}+|\mathbf p_{\mathbf r}|^2\right).
\end{equation}
Here $\!\rho_{\mathbf r}\!=\!\!\sum_\alpha n_{\alpha,\mathbf r}$ and $\!\mathbf p_{\mathbf r}\equiv\!\!(p_{1},p_{2},p_{3})_{\mathbf r}$ collects the three independent four-sublattice polarizations, with $p_{3,\mathbf r}=n_{A,\mathbf r}-n_{B,\mathbf r}-n_{C,\mathbf r}+n_{D,\mathbf r}$ defines $S_4$ in Eq.~(\ref{S4})~\footnote{$p_{1,\mathbf r}=n_{A,\mathbf r}+n_{B,\mathbf r}-n_{C,\mathbf r}-n_{D,\mathbf r}, p_{2,{\mathbf r}}=n_{A,\mathbf r}-n_{B,\mathbf r}+n_{C,\mathbf r}-n_{D,\mathbf r}$.}. At fixed $\rho_{\mathbf r}$, the interaction favors maximal intracell polarization, reached by a single-orbital texture. For the $A$-polarized top-left branch at the bifurcation $U_{c_1}$, $\psi_{B,\mathbf r}\simeq \psi_{C,\mathbf r}\simeq \psi_{D,\mathbf r}\simeq0$, the B and C components are correspondingly driven toward $Q(\beta_x)\equiv \gamma \psi_{A,\mathbf r}+\lambda \psi_{A,\mathbf r+\hat{x}}=0$ and $Q(\beta_y)\equiv\gamma \psi_{A,\mathbf r}+\lambda \psi_{A,\mathbf r+\hat{y}}=0$. These are the two directional BBH edge kernels. Their common normalizable solution is the top-left corner channel. Hence the interaction-selected internal texture exposes the corner kernel only when the carrier is compatible with both directional channels. For $\gamma>\lambda$, no decaying topological corner solution exists, leading to trivia corner excitation.

Figures~\ref{Fig4}(d) and (e) explicitly show the importance of bidirectional momentum matching for higher-order nonlinear correspondence. Keeping $k_{\min,x}=k_{{\rm top},x}=\pi$ while shifting $k_{\min,y}\neq k_{{\rm top},y}$ leaves the excitation localized but reduces $S_4$ and fidelity $\mathcal F_{\rm BBH}$ with the parent channel. 

\textcolor{blue}{\textit{Conclusion}} - We establish a qualitatively new regime of bulk-boundary correspondence in which topology remains predictively connected to emergent boundary structure after its conventional boundary signature is lost, without ever reopening the original band gap. This is possible when interactions, topology, and spin-degrees-of-freedom conspire to give rise to a compatibility principle. Our result is experimentally accessible in current atomic, polaritonic and photonic platforms~\cite{schafer_tools_2020,xie_topological_2020,wang_observation_2022,li_observation_2023,agrawal_two-dimensional_2024,st-jean_lasing_2017,solnyshkov_microcavity_2021,pernet_gap_2022,bisianov_stability_2019,guo_weakly_2020,mukherjee_observation_2020,maczewsky_nonlinearity-induced_2020,mukherjee_observation_2021,kirsch_nonlinear_2021,hu_nonlinear_2021,pernet_gap_2022,kartashov_observation_2022}. Engineered intercomponent or anisotropic couplings could favor different internal textures and expose rich boundary structures supported by the bulk. Extending this correspondence beyond the mean-field regime~\cite{heinsdorf_fate_2025} is an appealing future direction.

{\emph{Acknowledgment.}} This work was supported by the Key Project of the National Natural Science Foundation of China Joint Funds (Grants No.~U25A20197), and the National Natural Science Foundation of China (Grant No. 12374246). 

Cong Ding and Xuzhen Cao contributed equally.

%


\end{document}